\documentclass[sigconf]{acmart}
\usepackage{svg}
\usepackage{listings}
\usepackage{booktabs}
\usepackage{dirtytalk}  
\usepackage{makecell}
\usepackage{epstopdf}
\usepackage{graphics}

\usepackage{tikz}
\usepackage{graphicx}
\usepackage{rotating}
\usepackage[frozencache=true,cachedir=minted-cache]{minted}  

\usepackage{hyperref}
\usepackage{pgf-pie}  
\usepackage{listings}
\usepackage{xcolor}
\usepackage{media9}
\usepackage{lipsum}
\usepackage{pgfplots}

\usepackage{etoolbox}

\BeforeBeginEnvironment{figure}{\vskip-1ex}
\AfterEndEnvironment{figure}{\vskip-2ex}
\AfterEndEnvironment{table}{\vskip-5ex}

\definecolor{codegreen}{rgb}{0,0.6,0}
\definecolor{codegray}{rgb}{0.5,0.5,0.5}
\definecolor{codepurple}{rgb}{0.58,0,0.82}
\definecolor{backcolour}{rgb}{0.95,0.95,0.92}

\lstdefinestyle{mystyle}{
    backgroundcolor=\color{backcolour},   
    commentstyle=\color{codegreen},
    keywordstyle=\color{magenta},
    numberstyle=\tiny\color{codegray},
    stringstyle=\color{codepurple},
    basicstyle=\ttfamily\footnotesize,
    breakatwhitespace=false,         
    breaklines=true,                 
    captionpos=b,                    
    keepspaces=true,                 
    numbers=left,                    
    numbersep=5pt,                  
    showspaces=false,                
    showstringspaces=false,
    showtabs=false,                  
    tabsize=2
}
\usepackage{subfigure}
\usepackage{tcolorbox}
\usepackage{xcolor}
\usepackage{pdfpages}

\usetikzlibrary{shapes, arrows}
\usetikzlibrary{matrix, positioning, fit}

\newcommand{\todo}[1]{}
\newcommand{\uvp}[1]{}
\newcommand{\rh}[1]{}
\newcommand{\gal}[1]{}
\newcommand{\new}[1]{}
\newcommand{\old}[1]{}
\newcommand{\ourmod}[0]{\textsc{PragFormer}}
\newcommand{\ourdb}[0]{\textsc{Open-OMP}}
\newcommand{\compar}[0]{\textsc{ComPar}}
\newcommand{\lime}[0]{\textsc{LIME}}

\definecolor{dkgreen}{rgb}{0,0.6,0}
\definecolor{gray}{rgb}{0.5,0.5,0.5}
\definecolor{mauve}{rgb}{0.58,0,0.82}
\definecolor{darkblue}{rgb}{0.0,0.0,0.6}
\definecolor{cyan}{rgb}{0.0,0.6,0.6}
\definecolor{mBlue}{HTML}{4285f4}
\definecolor{mRed}{HTML}{ea4335}
\definecolor{mGreen}{HTML}{34a853}
\definecolor{mYellow}{HTML}{fbbc04}
\definecolor{mLightBlue}{HTML}{6d9eeb}
\definecolor{mLightRed}{HTML}{e06666}
\definecolor{mLightGreen}{HTML}{93c47d}
\definecolor{mLightYellow}{HTML}{ffd966}
\definecolor{archtBlue}{HTML}{9fc5e8}
\definecolor{archtTeal}{HTML}{a4d2d7}
\definecolor{archtYellow}{HTML}{ffe599}
\definecolor{archtOrange}{HTML}{f9cb9c}
\definecolor{archtCetus}{HTML}{ea9999}
\definecolor{archtOther}{HTML}{dd7e6b}
\definecolor{archtPurple}{HTML}{b4a7d6}
\definecolor{archtGray}{HTML}{eeeeee}
\definecolor{archtGreen}{HTML}{b6d7a8}
\definecolor{archtRed}{HTML}{ea9999}

\AtBeginDocument{%
  \providecommand\BibTeX{{%
    \normalfont B\kern-0.5em{\scshape i\kern-0.25em b}\kern-0.8em\TeX}}}

\setcopyright{none}

\acmConference[]{}{}{}

\begin{document}

\title{Learning to Parallelize in a Shared-Memory Environment with Transformers}
\author{Re'em Harel}
\email{reemha@bgu.ac.il}
\affiliation{%
  \institution{Department of Computer Science, Ben-Gurion University of the Negev}
    \institution{Department of Physics, Nuclear Research Center – Negev}
        \institution{Scientific Computing Center, Nuclear Research Center – Negev}
  \country{Israel}
}

\author{Yuval Pinter}
\email{uvp@cs.bgu.ac.il}
\affiliation{%
  \institution{Department of Computer Science, Ben-Gurion University of the Negev}
  \country{Israel}
}

\author{Gal Oren}
\email{galoren@cs.technion.ac.il}
\affiliation{%
  \institution{Department of Computer Science, Technion – Israel Institute of Technology}
    \institution{Scientific Computing Center, Nuclear Research Center – Negev}
  \country{Israel}
  }

\authornote{Corresponding author}

\begin{abstract}
  In past years, the world has switched to many-core and multi-core shared memory architectures.
  As a result, there is a growing need to utilize these architectures by introducing shared memory parallelization schemes to software applications.
  OpenMP is the most comprehensive API that implements such schemes, characterized by a readable interface.
  Nevertheless, introducing OpenMP into code, especially legacy code, is challenging due to pervasive pitfalls in management of parallel shared memory.
  To facilitate the performance of this task, many source-to-source (S2S) compilers have been created over the years, tasked with inserting OpenMP directives into code automatically.
  In addition to having limited robustness to their input format, these compilers still do not achieve satisfactory coverage and precision in locating parallelizable code and generating appropriate directives.
  In this work, we propose leveraging recent advances in machine learning techniques, specifically in natural language processing (NLP), to suggest the need for an OpenMP directive. The model also suggests directives complementary to S2S automatic parallelization compilers or provides immediate on-the-fly advice to the developer without compiling or executing the code.
  We create a database, \ourdb{}, specifically for this goal. \ourdb{} contains over 17,000 unique code snippets from different domains, half of which contain OpenMP directives while the other half do not need parallelization with high probability.
  We use the corpus to train systems to automatically classify code segments in need of parallelization, as well as suggest individual OpenMP clauses.
  We train several transformer models, named \ourmod{}, for these tasks and show that they outperform statistically-trained baselines and automatic S2S parallelization compilers in both classifying the overall need for an OpenMP directive and the introduction of \emph{private} and \emph{reduction} clauses.
  In the future, our corpus can be used for additional tasks, up to generating entire OpenMP directives.
  Our source code and database are available at: \textcolor{blue}{\url{https://github.com/pragformer/PragFormer}}.

\end{abstract}

\keywords{OpenMP, Shared memory Parallelism, Transformers, Source-to-source Compilers, Code Language Processing}

\begin{teaserfigure}
\centering
    \fontsize{6}{8} \selectfont 
        \scalebox{1}{\includegraphics{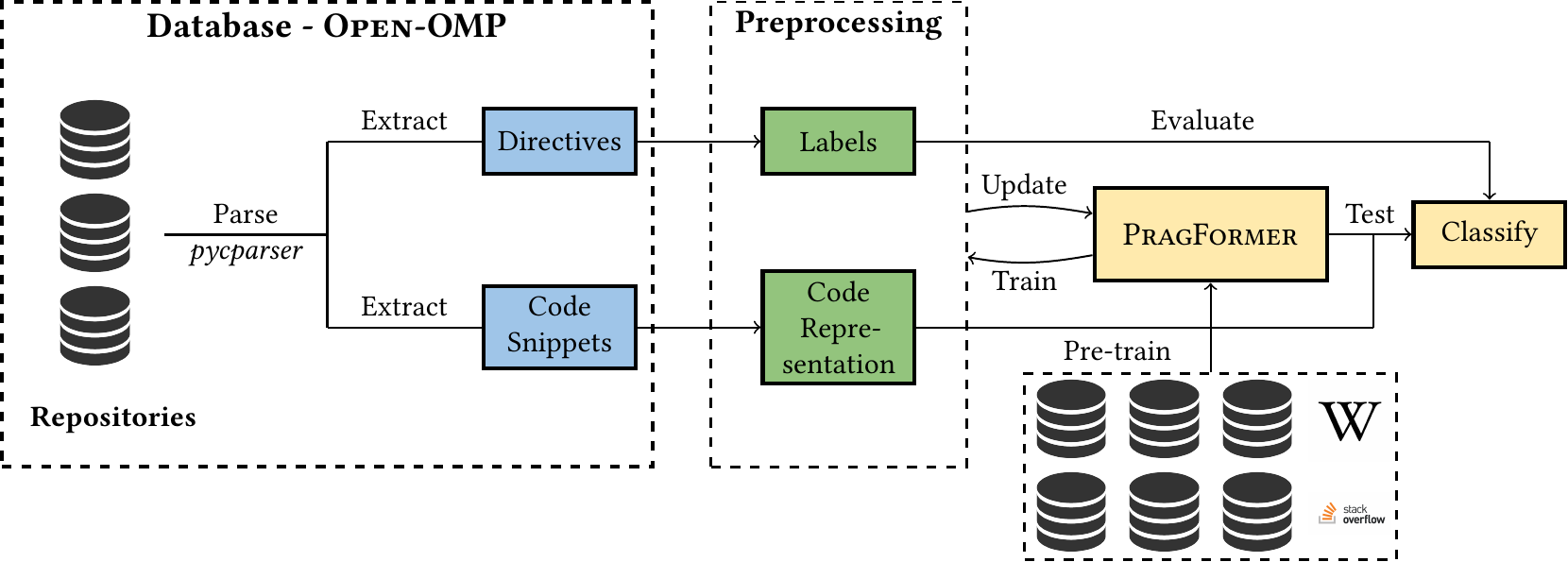}}
    \caption{Overview of the workflow for classifying OpenMP directives and clauses. \ourmod{} is our proposed model.}
    \label{fig:teaser}
\end{teaserfigure}
 \settopmatter{printfolios=true}

\maketitle

\section{Introduction}\label{sec:intro}

In past decades, software applications enjoyed a constant decrease in processor execution times, almost solely from the increase in performance of a single processor (based on Moore's law~\cite{brock2006understanding}).
However, due to the end of Dennard's scaling~\cite{dennard1974design}, the increase of a single processor performance (in clock time) has slowed down dramatically.
As a result of the constant demand for increased performance, the world has shifted its computational paradigm towards multi-core and many-core shared memory architectures (cache-coherent Non-Uniform Memory Access~\cite{lameter2013overview}, co-processors~\cite{reinders2012overview}, General Purposed GPUs~\cite{kirk2008nvidia}, among others).
This change yielded an increase in performance by utilizing shared memory parallelism schemes rather than increasing the performance of a single processor.
To adjust the trend, serial applications, as well as distributed ones, needed to be adapted to the new architecture and introduce shared memory parallelization schemes in order to exploit the new hardware.

The most comprehensive API that implements the shared memory model is the OpenMP API~\cite{dagum1998openmp}.
The OpenMP API consists of a set of compiler directives (\textit{pragmas}), library routines, and environment variables that allows a program to be executed in parallel (multi-threaded) within a shared memory environment.
One of the main advantages of the OpenMP API, which also contributes to its popularity, is its flexible and straightforward interface that is readable and easily interpreted.
For example, a \textit{private} clause states that each thread has its own private copy of a particular variable.
Nevertheless, introducing \textit{correct} and \textit{optimal} OpenMP parallelization instructions to applications is a complex and tedious task due to ubiquitous pitfalls in management of parallel shared memory, by architecture heterogeneity, and by the current necessity for human expertise to comprehend many fine details and abstract correlations~\cite{datadependency}.

\begin{table}[t]
  \begin{tabular}{l l l}
    \toprule
     \# & Code & S2S \\
     
    \midrule
1 & 
\begin{minipage}[t]{0.1\textwidth}
\begin{minted}{c}

for (i=0;i<=N;i++)
  A[i] = i;
  
for (i=0;i<=N;i++)
  B[i] = B[i]*2;
\end{minted}
\end{minipage}

 &
\begin{minipage}[t]{0.1\textwidth}
\begin{minted}{c}
#pragma omp parallel for 
for (i=0;i<=N;i++)
  A[i] = i;
#pragma omp parallel for
for (i=0;i<=N;i++)
  B[i]=B[i]*2;
\end{minted}
\end{minipage}

\\
\addlinespace

\hline
\addlinespace

2 &
\begin{minipage}[t]{0.2\textwidth}
\begin{minted}{c}
for (i=0;i<=N;i++) 
  if (MoreCalc(i))
     Calc(i);
\end{minted}
\end{minipage}
&
\begin{minipage}[t]{0.2\textwidth}
\begin{minted}{c}
#pragma omp parallel for 
for (i=0;i<=N;i++) 
  if (MoreCalc(i))
     Calc(i);
\end{minted}
\end{minipage}

\\

    \bottomrule
  \end{tabular}
    \caption{Pitfalls of S2S automatic paralellization compilers.}
  \label{table:code_snippet}
\end{table}
\begin{table}
  \begin{tabular}{l l}
    \toprule
     AST Representation \#1 & AST Representation \#2 \\
     
    \midrule
\begin{minipage}[t]{0.2\textwidth}
\begin{minted}{c}
For: 
  Assignment: =
    ID: i
    Constant: Int, 0
  BinaryOp: <=
    ID: i
    ID: N
  UnaryOp: p++
    ID: i
  Assignment: =
    ArrayRef: 
      ID: A
      ID: i
    ID: i
For: 
  Assignment: =
    ID: i
    ...
\end{minted}
\end{minipage}
& 
\begin{minipage}[t]{0.2\textwidth}
\begin{minted}{c}
For: 
  Assignment: =
    ID: i
    Constant: Int, 0
  BinaryOp: <=
    ID: i
    ID: N
  UnaryOp: p++
    ID: i
  If: 
    FuncCall: 
      ID: MoreCalc
      ExprList: 
        ID: i
    FuncCall: 
      ID: Calc
      ExprList: 
        ID: i
\end{minted}
\end{minipage}
\\

    \bottomrule
  \end{tabular}
    \caption{Corresponding AST representations of the examples in \autoref{table:code_snippet}.}
  \label{table:ast_snippet}
\end{table}

\vspace{10ex}
\subsection{Source-to-Source Automatic Parallelization Compilers}
Over the last decade, automatic parallelization compilers have been created to ease the process of introducing parallelization directives into code~\cite{yao2016survey, prema2019study, prema2013analysis, blair2012parallel}. 
Generally, the automatic parallelization process occurs during compilation time or via source-to-source (S2S) automatic parallelization compilers that insert OpenMP directives automatically.
The major drawback of automatic parallelization during the compilation process is the nontransparent result that they generate, i.e., it is unknown if and how automatic parallelization was achieved.
In contrast, S2S compilers provide the full output in the source code, providing the user with full transparency of the result, and allowing the user to review, commit changes and even optimize the output~\cite{dipasquale2005comparative}.
Examples of such S2S compilers are: Par4All~\cite{par4allhome}, AutoPar~\cite{quinlan2000rose}, and Cetus~\cite{cetushome}.
Generally, the workflow of said S2S compilers is as follows:
\begin{enumerate}
    \item Create an abstract syntax tree (AST)~\cite{neamtiu2005understanding} using a source code parser, such as ANother Tool for Language Recognition (ANTLR)~\cite{parr2013definitive}.
    \item Apply data dependence algorithms~\cite{fagin1984theory} (predefined rules) on the result of (1).
    \item Produce the appropriate OpenMP directive~\cite{kennedy2001optimizing} based on (2).
\end{enumerate}

Harel et al.~\cite{harel2020source} (2020) and S. Prema et al.~\cite{prema2017identifying, prema2019study} (2017, 2019) showed that although S2S compilers provide a solution for code parallelization, they still have many pitfalls.
Among these pitfalls, S2S compilers produce suboptimal directives (as judged by human experts), degrade performance, and sometimes even fail to insert a directive.
For example, in example \#1 in \autoref{table:code_snippet}, the original code contains two independent arrays and two loop segments: initialization of array~\textit{A} and a simple calculation of array~\textit{B}. 
The S2S compilers can produce OpenMP directives that create unnecessary overhead caused by spawning the threads twice, once for each loop, rather than just once (the compilers cannot comprehend such a case as it requires additional knowledge regarding the internal logic of the code).
This overhead can easily be avoided by identifying several consecutive loop segments, wrapping them with a single parallel region directive, and even adding the \textit{nowait} directive---if the consecutive loop segments are independent---to further improve performance. 
Furthermore, sometimes a \textit{for}-loop exhibits an unbalanced workload where each iteration takes a different amount of time to finish, such as that created by the \textit{if} statement in example \#2.
In such cases, S2S compilers will not make use of the \textit{schedule(dynamic)} directive, which distributes the workload between the threads dynamically, and instead use the default \textit{schedule(static)}, producing a non-optimal increase in performance. 
These mistakes are caused by the inability of the deterministic nature of the S2S compilers to identify loops with an unbalanced workload from the source code alone.
Identifying the unbalanced workload usually requires understanding the internal logic of the code, or identifying that the \textit{if} statement contains extensive calculations compared to the \textit{else} statement.

The examples detailed above are not exceptional; it was found in~\cite{harel2020source} that determining function side effects is a significant issue for these S2S compilers.
In addition, applying the data dependence algorithm on the AST representation of the source code consumes significant time and memory dependent on the number of lines inside the loop's scope, which makes these S2S compilers impractical for long code segments.
As an example, in \autoref{table:ast_snippet} we present the AST representations of the code in \autoref{table:code_snippet}.
The AST, which has a tree structure, is created by parsing the syntax of the source code, with each line containing a single token corresponding to a single operation or variable.
For example, the line \textit{Assignment} corresponds to the operator \textit{=}. 

Nevertheless, S2S compilers applicability in the HPC community is negatively perceived~\cite{milewicz2021negative}.
The following four points were presented by Milewicz et al.~\cite{milewicz2021negative} (2021) as the main issues with S2S transformation:
\begin{enumerate}
    \item S2S transformations interfere with or are oblivious to downstream compiler optimizations.
    \item S2S compilers are difficult to extend to support new programming models.
    \item S2S approaches lead to complex and fragile workflows.
    \item Parsing source code is inherently difficult, as it requires parsing the source code, representing the variables and dependencies, and then applying some algorithm over the representation. Nevertheless, the problem is inherently difficult as representation is not comprehension nor intelligence.
\end{enumerate}
Given this perception, there are problems that are difficult to solve by classical algorithms (such as existing S2S compilers).
Instead, solving these problems---which are similar to problems in language processing---by leveraging machine learning techniques, specifically natural language processing (NLP) techniques, can be useful. For example, with NLP techniques, the parsing stage can be almost wholly skipped as the model's input can be the raw text of the source code.

\subsection{Code Language Processing}
Natural language processing is a field in artificial intelligence concerned with giving computers the ability to process and analyze natural language text, mostly through technologies from machine learning (ML)~\cite{jurafsky2019speech}.
Popular tasks in NLP include machine translation, question answering, information extraction, and named entity recognition~\cite{eisenstein2019natural}.
Due to some similarities between natural languages and programming languages, many studies have explored the prospects of applying NLP methods to code.
These applications include source code analysis~\cite{sun2014empirical,imam2020use, sharma2021survey}, where code is processed to generate natural-language documentation~\cite{richardson-etal-2017-code2text,leclair2020improved,haque2020improved} or further code; source code generation~\cite{yin-neubig-2017-syntactic,parvez-etal-2021-retrieval-augmented} based on natural language descriptions; and lighter applications where no generation happens, such as programming language classification~\cite{ugurel2002s} and code change analysis~\cite{tollin2017change}.
Methods proposed for these tasks are often rooted in the development of cognitive models for human comprehension of code~\cite{program_comprehension_iwpc,BENNETT20021}.
Together, these methods and tasks can be grouped together under the umbrella term Code Language Processing (CLP).

The applicability of NLP tools, particularly the most effective ones that use ML and its recent development deep learning (DL), to tasks applying to code languages is not at all trivial.
Code is firmly structured, unambiguous, and purposeful, as opposed to natural languages which exhibit many \say{noisy} characteristics, as well as cultural effects, ubiquitous ambiguity, and varied levels of informational purpose~\cite{nguyen2016computational}.
Many NLP tasks are also characterized by being cast locally, on a single-sentence level (such as sentence-level translation), an assumption which is starting to be challenged and abandoned~\cite{dai-etal-2019-transformer,beltagy2020longformer}; in the code landscape, long-range dependencies are almost unavoidable, considering the scopes over which variables are used and functions are called.
Therefore, use of powerful but local sequential models such as the recurrent neural net (RNN)~\cite{elman1991distributed} or its more robust variant the long short-term memory network (LSTM)~\cite{hochreiter1997long} may not produce good results for the tasks at hand.
At the present time, the most used ML technique for complicated CLP problems (such as program comprehension) is the attention-based encoder-decoder~\cite{DBLP:journals/corr/LuongPM15,sharma2021survey} model known as the transformer~\cite{DBLP:transformers}.
In this paper, we present a transformer-based method for facilitating code parallelization through OpenMP, demonstrating its improved performance over S2S systems, as well as the necessity of its sophisticated model architecture.
\section{Code Language Parallel Processing}\label{sec:clpp}
To better formulate the above proposed idea, we hereby introduce code language parallel processing (CLPP) as a sub-discipline in CLP which deals with incorporating parallelism schemes via ML techniques.
While ML techniques are not strange to the parallel processing area~\cite{chen2009adaptive,woodsend2009hybrid,milani2020autotuning,mishra2020using,wang2020machine,kurth2016openmp, hernandez2018using, wang2009mapping}, as of yet there has not been an attempt to use CLP techniques to insert OpenMP directives automatically. Similar to the program comprehension task, inserting OpenMP directives automatically requires comprehending the data dependencies and logic of the code in order to produce correct and optimal OpenMP directives. Nevertheless, given that the amount of existing OpenMP directives is limited, and in practice, there are only dozens of OpenMP directives that are widely being used in the HPC community~\cite{mattson2019openmp}; this task can be categorized as a classification rather than generation task. 

We note that OpenMP directives syntax is closely related to documentation as they both are written as comments in the code; documentation appears as comments for the developer, and OpenMP directives as comments for the compiler. 

As stated by Milewicz et al.~\cite{milewicz2021negative}, parsing the source code to an AST, or other representation, is the first and most crucial step in every S2S solution. However, HPC practitioners believe that this step is far from being solved and is the main reason why S2S compilers are negatively perceived~\cite{milewicz2021negative}. 
Nonetheless, due to the recent innovations in ML, specifically in CLP~\cite{sharma2021survey}, and widely available source code databases, the possibility of replacing the AST step and suggesting code segments that can benefit from OpenMP directives with newer techniques rises.

\subsection{Research Objectives}\label{sec:openmp_directive_classification}
In order to evaluate and explore the possibility of replacing the standard S2S compilers with NLP models, we formulate two research questions and focus our work on answering them.

\subsubsection{RQ1: Is this Code Parallelizable?}
First, we wish to identify whether a given code segment is in need of an OpenMP directive.
As stated in~\cite{harel2020source}, one of the pitfalls of S2S compilers, and of introducing OpenMP directives generally, is the unnecessary generation of OpenMP directives in cases where parallelization is counterproductive.
For example, if a for-loop construct contains a small amount of computer-operations (such as a low iteration count), the overhead of spawning OpenMP threads outweighs the gain to be made by parallelizing the construct.
However, in some cases such as array initialization, introducing OpenMP directives might be beneficial in speeding up other OpenMP directives, due to the first-touch policy in cc-NUMA~\cite{lameter2013overview}.

\subsubsection{RQ2: Does this Parallelizable Code Need a Certain Clause?}
One of the problems in executing a program concurrently is possible data races, which under some circumstances cause the code to be inconsistent across separate runs.
In some cases, these problems can be eliminated by adding specific OpenMP clauses~\cite{oren2017automp}, such as the \textit{private} clause and the \textit{reduction} clause.
Thus, we define an additional task of classifying the need for a \textit{private} or a \textit{reduction} clause affecting any of the variables in a given code segment which we already know to be amenable to parallelization.

Together, these tasks complement each other and can be seen as the beginning of creating a full pipeline which generates OpenMP directives automatically.
In the meantime, due to the negligible inference time (contrary to S2S compilers), it can be used as an immediate \say{advisor} for developers to identify locations that can benefit from an OpenMP directive---even with partial or on-going programming, which is impossible with current S2S compilers. Moreover, the model and the S2S compilers can be incorporated such that in cases both the model and the S2S compilers agree on a directive, it will remain. Thus, verifying the correctness of the directive and the necessity.

\section{Data}\label{sec:data}
\subsection{Corpus}
In order to train and evaluate models on the task of code parallelization, we created a database, or corpus, of code files, which we call \ourdb{}\footnote{\textcolor{blue}{\url{https://github.com/pragformer/PragFormer/blob/main/Open_OMP.tar.gz}}}.
We queried \emph{\url{github.com}}, the most well-known code repository housing website, via a script named \textit{github-clone-all}\footnote{\textcolor{blue}{\url{http://github.com/rhysd/github-clone-all}}}, which allows searching for source files by repository.
We extracted only C files from repositories containing the phrase \say{OpenMP} in the title, description or \textit{README} file.
This query resulted in more than 20,000 files from approximately 7,000 different repositories.

\subsubsection{Inclusion Criteria}
Among the 20,000 files, we considered only files written in C and containing an OpenMP directive.
We used \emph{pycparser}~\cite{bendersky2010pycparser}, a Python based C code parser, to identify these files.
The parser creates an AST of the source code, and by traversing the tree, one can easily identify OpenMP directives along with their corresponding loop segments. 
We identified the question of whether or not a code segment is even suitable for OpenMP annotation as an important task for a model to be able to answer (\textit{RQ1}), which means that the corpus must also include code that is not explicitly parallelizable in order to provide ML classifiers with negative examples to learn from.
We limited our negative data to code without OpenMP directives appearing in files where elsewhere such directives do exist, to rule out cases where code amenable to directives was not annotated due to developers unfamiliar with parallelization schemes or those who work with incompatible hardware.

\subsubsection{Exclusion Criteria}
Some OpenMP directives are nearly impossible to predict automatically from code alone; for example, the \emph{task} construct requires extensive background knowledge regarding the logic of the code. Thus, in accordance with S2S compilers, we include only OpenMP directives defined over a loop segment (directives containing \emph{\#pragma omp parallel for}) in the corpus. 
In addition, we found numerous repositories containing examples of OpenMP directives that were created solely for testing compiler compatibility.
These examples, which contained an empty for-loop, were also excluded from the corpus.
In the developer community, it is common to copy source code from other existing projects. Therefore, to verify that each entry is unique, it was scanned for similar entries and replicas were removed.

\begin{figure}[H]
\centering
    \fontsize{6}{8} \selectfont 
        \scalebox{0.65}{\includegraphics{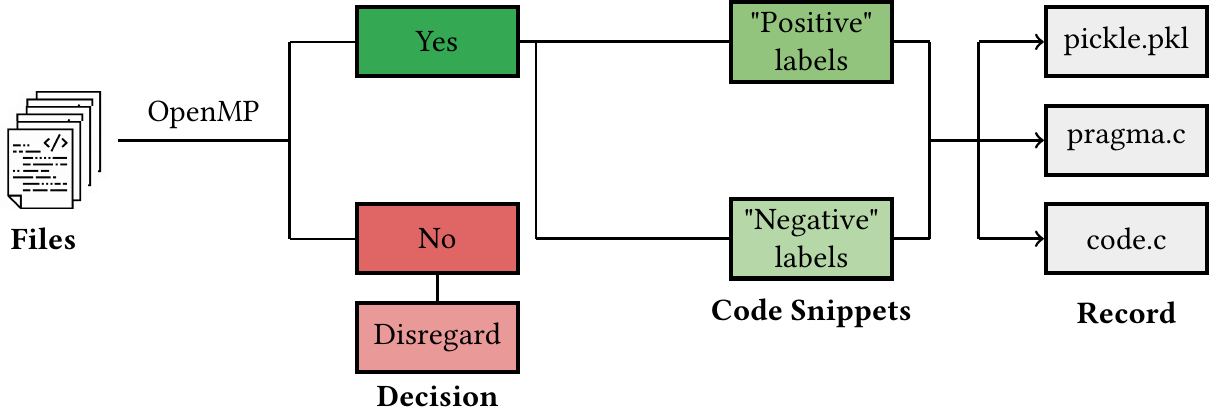}}
    \caption{Overview of the workflow for creating the database.}
    \label{fig:database_workflow}
\end{figure}

The corpus thus contains two types of records: those that contain examples with OpenMP directives, and those that do not.
Each record contains three files: (1) the code segment relevant to the directive---the loop segment along with, if found, implementations of functions called inside the loop segment; (2) the OpenMP directive; and (3) a pickle file containing the code segment's and the OpenMP directive's AST (generated from \emph{pycparser}). In most cases, the context of the loop segment is sufficient to produce an OpenMP directive. Moreover, in order to create adequate labeling, we needed to create a clear link between the feature (code) and label (directive). Therefore, the database currently includes the OpenMP directive and its corresponding loop.
\autoref{fig:database_workflow} summarizes the creation process of the corpus.

\begin{figure}[ht]
\centering
\begin{tikzpicture}[scale=0.650]
\pie[color={archtOrange,archtBlue,archtRed,mGreen}]{
    33.5/Unknown (no README),
    16.5/Benchmark,
    7/Testing,
    43/Generic Application}
\end{tikzpicture}
\caption{The distribution of OpenMP snippets sources.} 
\label{fig:pie_chart_readme}
\end{figure}
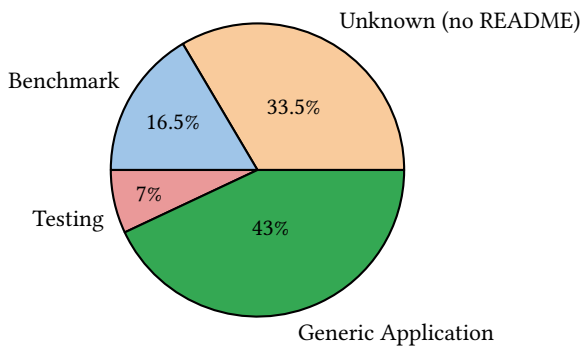

We present statistics of our database by: (1) the different OpenMP directives count (\autoref{table:commonopenmp}), (2) the number of lines count of the code snippets (\autoref{table:number_lines_database}), and (3) the domain distribution of the code snippets (\autoref{fig:pie_chart_readme}): if the code snippet doesn't contain a README file (thus, unknown domain), the README file contains the keyword 'benchmark' or 'testing' and the default case which is assumed to be a generic application. As can be seen, most of the database contains small to medium code snippets, with diverse OpenMP directives, and mostly from real applications.\footnote{Full github repositories url's by domain distribution: \textcolor{blue}{ \url{https://github.com/pragformer/PragFormer/blob/main/log_repositories_readme.txt}}}

\begin{table}[H]
    \begin{tabular}{lr}
    \toprule
    \textbf{Description}  & \textbf{Amount}  \\
    \midrule
    Total code snippets & 17,013
    \\
    For loops with OpenMP directives & 7,630 
    \\
    Schedule \textit{static} & 7,256
    \\
    Schedule \textit{dynamic} & 374
    \\
    Reduction & 1,455
    \\
    Private & 3,403
    \\
    \bottomrule
    \end{tabular}
    \caption{Statistics of the OpenMP directives on the raw database.}
    \label{table:commonopenmp}
\end{table}

\begin{table}[H]
    \begin{tabular}{lr}
    \toprule
    \textbf{Line Count}  & \textbf{Amount}  \\
    \midrule
    $<$ 10 & 9,865
    \\
    11--50 & 5,824
    \\
    51--100 & 724
    \\
    $>$ 100 & 600
    \\
    \bottomrule
    \end{tabular} 
    \caption{Code snippet lengths in the raw database.}
    \label{table:number_lines_database}
\end{table}

\subsection{Datasets}\label{sec:split}

An essential component of any ML model built for a supervised task such as ours is the \textit{dataset}, a group of \textlangle{}\textit{input}, \textit{label}\textrangle{} instances split into three sets---training, validation, and test.
The training set contains examples used for training the model (updating internal parameter values); the validation (sometimes called development) set contains examples used for making model-level decisions that cannot be tuned per instance (termed hyperparameters) and helps avoid over-fitting the model being trained on the specific idiosyncrasies of training set examples; and the test set contains examples used to evaluate the effectiveness of the overall resulting model against other systems.
We thus divided our data into the required splits using a 80\%--10\%--10\% ratio.

We created two different datasets for our task, one for each research question (or task).
The first dataset contains code samples from the database which have (or don't have) corresponding OpenMP directives, labeled positive (or negative, respectively), and is used to decide whether an OpenMP directive is needed for a given code snippet.
The second dataset contains only instances known to have OpenMP directives in the database, and is used for the task of classifying \textit{private} and \textit{reduction} OpenMP clauses.
In \autoref{table:dataset} we present the dataset sizes. 
The train/validation/test splits in both datasets were performed randomly at the instance level, while maintaining a balanced positive-negative label distribution in each dataset.

\begin{table}[H]
\begin{tabular}{lrr}
\toprule
   \textbf{Dataset}  & \textbf{Directive} & \textbf{Clause}\\ 
\midrule
Training & 14,442 & 6,482
 \\
Validation & 1,274 & 572
 \\
Test & 1,274 & 572
 \\
\bottomrule
\end{tabular} 
\caption{Amount of examples in each dataset for the OpenMP directive and clause classification.}
\label{table:dataset}
\end{table}

\section{\ourmod}
We propose a novel model, \ourmod, for identifying OpenMP directives and clauses.
\ourmod{} is based on the transformer architecture~\cite{DBLP:transformers}, and the workflow of applying it is presented in \autoref{fig:teaser}.
This pipeline resembles the workflow of program comprehension models~\cite{sharma2021survey} and of S2S compilers:
(1) tokenize the source code in order to allow its ingestion by the transformer model (\S\ref{sec:code_representation}); (2) train the transformer-based model (\S\ref{sec:model}) to classify OpenMP directives from the data segments in the corpus; (3) and evaluate the predicted OpenMP directives, as described in~\S\ref{sec:results}. 
In this work, we train models for three separate tasks: OpenMP directive identification; \textit{private} clause identification; and \textit{reduction} clause identification.

The flexible form of machine learning systems allows us to initialize the same model architecture for all three tasks, which we cast as classification problems (whether or not a directive/clause is needed), letting only the values of internal numeric parameters to diverge between the individual task models through backpropagation as the training procedure progresses.

\begin{table*}
  \begin{tabular}{lm{14cm} lm{7cm}}
    \toprule
     Representation & Example \\
     
    \midrule
Text & 
for (i = 0; i < len; i++)  a[i] = i;
\\
\addlinespace

Replaced-Text & 
for (var0 = 0; var0 < var1; var0++) arr0[var0] = var0;
\\
\addlinespace

AST & 
For: Assignment: = ID: i Constant: int, 0 BinaryOp: < ID: i ID: len UnaryOp: p++ ID: i Assignment: = ArrayRef: ID: a ID: i ID: i
\\
\addlinespace

Replaced-AST & 
For: Assignment: = ID: var0 Constant: int, 0 BinaryOp: < ID: var0 ID: var1 UnaryOp: p++ ID: var0  Assignment: = ArrayRef: ID: arr0 ID: var0 ID: var0

\\
    \bottomrule
  \end{tabular}
    \caption{Examples of the different code representations considered.}
  \label{table:code_representation}

\end{table*}

\subsection{Model}\label{sec:model}
\ourmod{} is composed of a transformer model followed by a fully-connected (FC) layer which performs classification. 
The transformer architecture~\cite{DBLP:transformers} has gained popularity due to its impressive ability to leverage data from vast unlabeled sequence resources in a \textit{pre-training} phase and apply it to a large range of end tasks, a property known as \textit{transfer learning}.
At a high level, the transformer is a model which receives a sequence of vectors as its input and passes it through a massively-parameterized \textit{encoder} module which outputs a sequence of \textit{contextualized} vectors of the same length.
These, in turn, are fed into a \textit{decoder} which can either generate a sequence of symbols from a pre-set vocabulary (as in the case of, for example, machine translation), or perform a classification task such as the one at hand (or, for example, named entity recognition).

The encoder, in turn, is made up of identical \textit{layers} of parameterized architectural \textit{blocks}, each producing its own sequence of representation vectors.
In theory, output vectors become more and more context-dependent after each layer's treatment.
The main mechanism allowing this behaviour is the \textit{self-attention} component within the blocks.
This component calculates a numeric score for each position in the input sequence with respect to the other ones, based on their input vectors and on parameterized projection matrices.
The score then determines the level of consideration in preparing this position's output for ingestion by the next layer.
For example, in a passage over the English sentence \say{The house does not have a well}, a self-attention block may compute a high score in the position corresponding to `well' with respect to that of `house', resulting in an output representation more approximating the \say{dug-out structure} sense of `well' as opposed to that of the \say{in a good way} sense (and others).
In a decoder, the attention mechanism is also applied to items from the input, enabling \textit{cross-attention}~\cite{bahdanau2014neural}.
We refer the reader to~\cite{DBLP:transformers} for more details.

In the realm of source code, attention scores may represent how much influence each of the other variables or statements have over a given input variable's contextualized vector.
This enables even elements which are far away, position-wise, to affect any output; a typical transformer model's sequence length is capped at hundreds of tokens. 

We implement \ourmod{} based on the pre-trained state-of-the-art DeepSCC~\cite{DBLP:deepscc} model. Similarly to our task, which is identifying and advising the need for an OpenMP directive, DeepSCC was trained with the aim of classifying the programming language of a source code. In DeepSCC the self-attention mechanism helps focusing on specific scopes (of variables or operations) to determine the program language. This mechanism is crucial in our task, as the identification of the need of an OpenMP directive is hinted in long-range dependencies in the source code, which the self-attention mechanism of the transformer model is most suited to. 
As part of our research focus, we want to examine whether this code-focused task provides adequate initialization for training our model on our parallelization tasks.
As such, it can be seen as an instance of transfer learning into a \textit{low-resource} scenario, a technique shown to be effective in many applications of transformers in NLP~\cite{zoph-etal-2016-transfer,han-eisenstein-2019-unsupervised,imankulova-etal-2019-exploiting,kasai-etal-2019-low,sarioglu-kayi-etal-2020-detecting}.
DeepSCC was created by fine-tuning a pre-trained RoBERTa~\cite{liu2019roberta} model on a large corpus of source code files (including C and C++, which approximate our corpus) via the masked language model (MLM) objective~\cite{taylor1953cloze}.
In MLM, a random subset of tokens in the input are obscured from the encoder (\say{masked}), and in a self-supervised manner, the model tries to predict the masked tokens based on the encoder's output, by selecting it from its vocabulary. Moreover, the attention mechanism in DeepSCC learns helps the model emphasize specific 
Ultimately, DeepSCC manages to fine-tune the parameters of the RoBERTa model, which were originally trained to process English text, to be better suited for source code as well as suppressing the results of other models tasked with the same objective (achieving 87\% accuracy). Thus, we believe DeppSCC provides adequate initialization and providing an apt starting point for \ourmod{}.

Following the encoder's operation over source code input, the output vectors are fed into a FC layer which predicts a binary label based on the individual task (identification of a need for a directive, or for of a specific clause) through a softmax layer which transforms the real-valued scores output from the FC layer into probabilities.
If the probability is computed to be over 0.5, \ourmod{} predicts a positive outcome.
During training, a \textit{cross-entropy loss} is computed based on the predicted probability of the correct label, and this value is backpropagated through the entire network (including the encoder layers and the input vectors, also known as \textit{embeddings}) to update its parameters.
The binary cross-entropy loss for a single input instance $x$ with the (true) label $y\in\{0,1\}$ is given by:
\begin{equation}\label{eq:cross_entropy}
\mathcal{L}_{BCE}(x,y)=-(y\log(p(x))+(1-y)\log(1-p(x))),
\end{equation}
where $p$ is the probability assigned by the final softmax layer to the positive label.

\subsection{Representing the Code} \label{sec:code_representation}
Raw source code cannot serve as an input to an ML model; instead, transforming the code to a sequence of tokens from a pre-set vocabulary is needed.
This transformation process, known as \emph{tokenization}, is then followed by associating each keyed token with a numerical vector (\textit{embedding}) which is used as input to the model, to be modified along with the rest of the model's parameters during training.

One possible way to tokenize source code is as if it were text.
This way, each keyword, operator, identifier and symbol would be assigned its own token~\cite{sharma2021survey}. 
However, due to the structured nature of code, previous work~\cite{sun2019grammar, jayasundara2019treecaps} suggests that tokenizing the abstract representation of the code (such as AST or CDG) rather than the raw code itself can be helpful in order to perform source code analysis. 
Representing the code this way has been shown to improve results in various code understanding tasks compared to a lexical representation~\cite{hu2018deep,leclair2020improved, DBLP:journals/corr/abs-2004-04881}.
Another line of work proposes a combination of AST and raw code, either by embedding them both separately and combining their vectors inside the model, or by mixing them directly at the textual level, finding it to be beneficial over using each kind separately~\cite{hu2020deep,li2022setransformer}. 

Although other source code representations achieve proper word embeddings, such as \emph{word2vec}~\cite{church2017word2vec}, \emph{IR2vec}~\cite{venkatakeerthy2020ir2vec} and \emph{Programl}~\cite{cummins2020programl}, the tokenizer and the pre-trained model (DeepSCC and RoBERTa) were trained exclusively on natural language text. Therefore, exploiting the transfer learning property with different representations other than text will likely produce poor results (as exemplified with the AST representation in this work). Overcoming this limit can be achieved by training the tokenizer and the base model (i.e., RoBERTa) from scratch with a large corpus of the appropriate representation.
Therefore, we argue that it might be easier for \ourmod{} to understand the dependencies from raw source code, rather than from its AST representation.
We consider both approaches---AST representation and raw code (`text') representation---as input for the model's tokenizer.

A frequent problem in applying pre-set vocabulary tokenizers to any textual input is the representation of strings not present in the vocabulary, known as out-of-vocabulary items, or \textit{OOVs}~\cite{brill-1995-transformation}.
When encountered, OOVs are either broken down into manageable units, even at the character level, or mapped to some default catch-all token (often styled \emph{<UNK>} for \say{unknown}) which encumbers learning useful representations for their semantics.
In the code domain, the OOV problem can manifest itself mainly through identifier naming conventions, which vary greatly across organizations and communities, with many developers opting for their own idiosyncratic names for variables and functions.
This has been shown to harm model performance in code understanding tasks~\cite{hellendoorn2017deep, hu2020deep, hu2018deep}.
While most previous work~\cite{hellendoorn2017deep, hu2020deep, hu2018deep,leclair2020improved} has opted for the \textit{<UNK>} solution, classifying OpenMP directives does not require knowledge of individual identifier names, and we thus replace them during a pre-processing step with predefined indexed words such as \textit{var1}, \textit{var2}, reducing the required vocabulary size for the model's tokenizer and avoiding many OOVs.
We hypothesize that this step might also provide the effect of learning more meaningful representations for our canonical name tokens, since they are now shared across training instances.

Altogether, we experiment over four different code representations for our model: text, text with replaced identifiers (\textit{R-Text}), AST, and AST with replaced identifiers (\textit{R-AST}).
As the AST has a tree structure, it cannot be fed directly to a model that expects a sequence as an input.
Therefore, we obtain an adequate representation by applying the DFS algorithm to the AST. 

An example of each of the four representations is shown in \autoref{table:code_representation}. In \autoref{table:database_tokens} we present statistics at the token level for our dataset given each of the pre-tokenization schemes.
Note the distinction between token counts (individual symbols, including duplicates) and symbol types (unique symbols, an entry in a dictionary/vocabulary)~\cite{maciver1937token}.
In terms of current transformer models, such vocabulary sizes and OOV token (or type) counts as the ones present in all representations are relatively small, suggesting that the raw textual representation might be sufficient for obtaining good results.
In addition, the average number of tokens per code snippet is lower for text representation, since the AST representation adds words used to describe operational logic.

\begin{table}[H]
\begin{tabular}{lrrrr}
\toprule
   & \textbf{Text}  & \textbf{R-Text} & \textbf{AST} & \textbf{R-AST}  \\ 
\midrule
Train vocab size & 6,427 & 2,424 & 5,261 & 3,409
 \\
OOV types & 398 & 226 & 348 & 309
\\
Avg. length & 33 & 30 & 37 & 35
\\
\bottomrule
\end{tabular} 
\caption{Type-level corpus statistics. `OOV types' refers to the number of symbol types in the validation and test sets which are not found in the training set. `Avg. length' is the average amount of tokens in a code segment (for-loop).}
\label{table:database_tokens}
\end{table}

\subsection{Implementation}
In \ourmod, we use the tokenizer from DeepSCC-RoBERTa.\footnote{\textcolor{blue}{\url{https://huggingface.co/NTUYG/DeepSCC-RoBERTa}}}
The largest token length in a code snippet was 110, and so it was set as the maximum length for the input to the model.
We fine-tuned the DeepSCC model in \ourmod{} to our dataset, meaning the internal parameters already pre-trained on the language modeling task were also updated during our model's training phase.
The FC layer in \ourmod{} contains two dense layers with a ReLU activation function between them.
We implemented dropout as a regularization strategy to avoid overfitting the training set~\cite{srivastava2014dropout}.
We updated model parameters using backpropagation, implemented via the AdamW gradient descent optimizer~\cite{loshchilov2018fixing}.

\section{Results}\label{sec:results}
To evaluate the effectiveness of \ourmod, we conduct several experiments over our extracted dataset (\S\ref{sec:data}).
We first determine the best code representation for the tasks (cf. \S\ref{sec:code_representation}), then follow up by comparing the best-represented \ourmod{} variant against other models, namely an S2S compiler and a standard ML model.


\subsection{Code Representation}
\label{sec:validres}
A crucial step in performing source code analysis with ML models is representing the code so that the model can extract the maximum information from it and infer upon it.
We compare the performance of models trained on each of the four code representations presented in~\S\ref{sec:code_representation}, namely: text, replaced text, AST, and replaced AST.
The training process was conducted on the OpenMP directive classification dataset presented in~\S\ref{sec:split}, and we calculated the accuracy of each trained model on the validation set.

\vspace{9ex}
\begin{figure}[htbp!]
\includegraphics*[width=7.5cm]{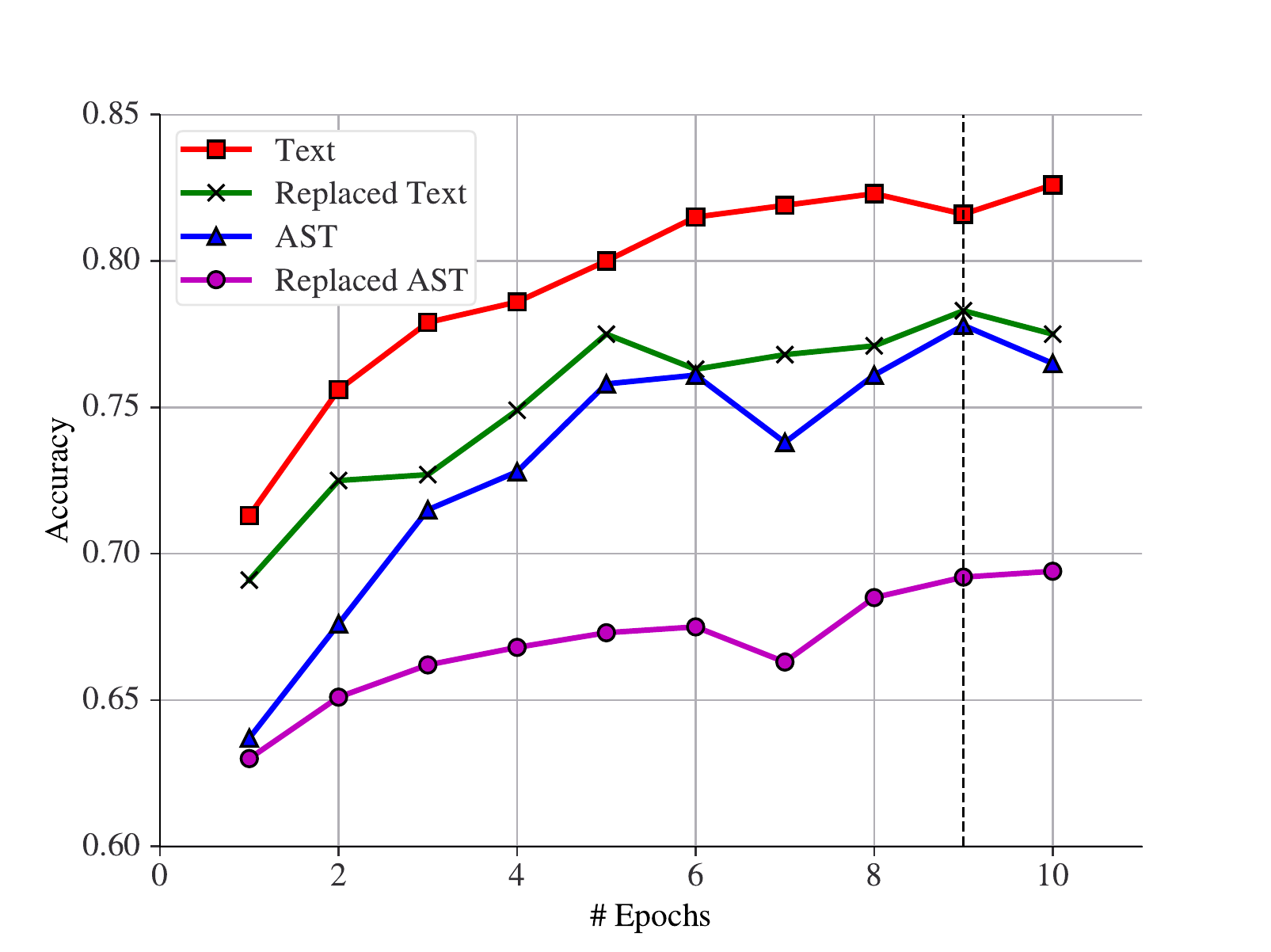}
\caption{Accuracy of the four source code representations as training epochs increase.}
\label{fig:accuracy}
\end{figure}

\begin{figure}[htbp!]
\includegraphics*[width=7.5cm]{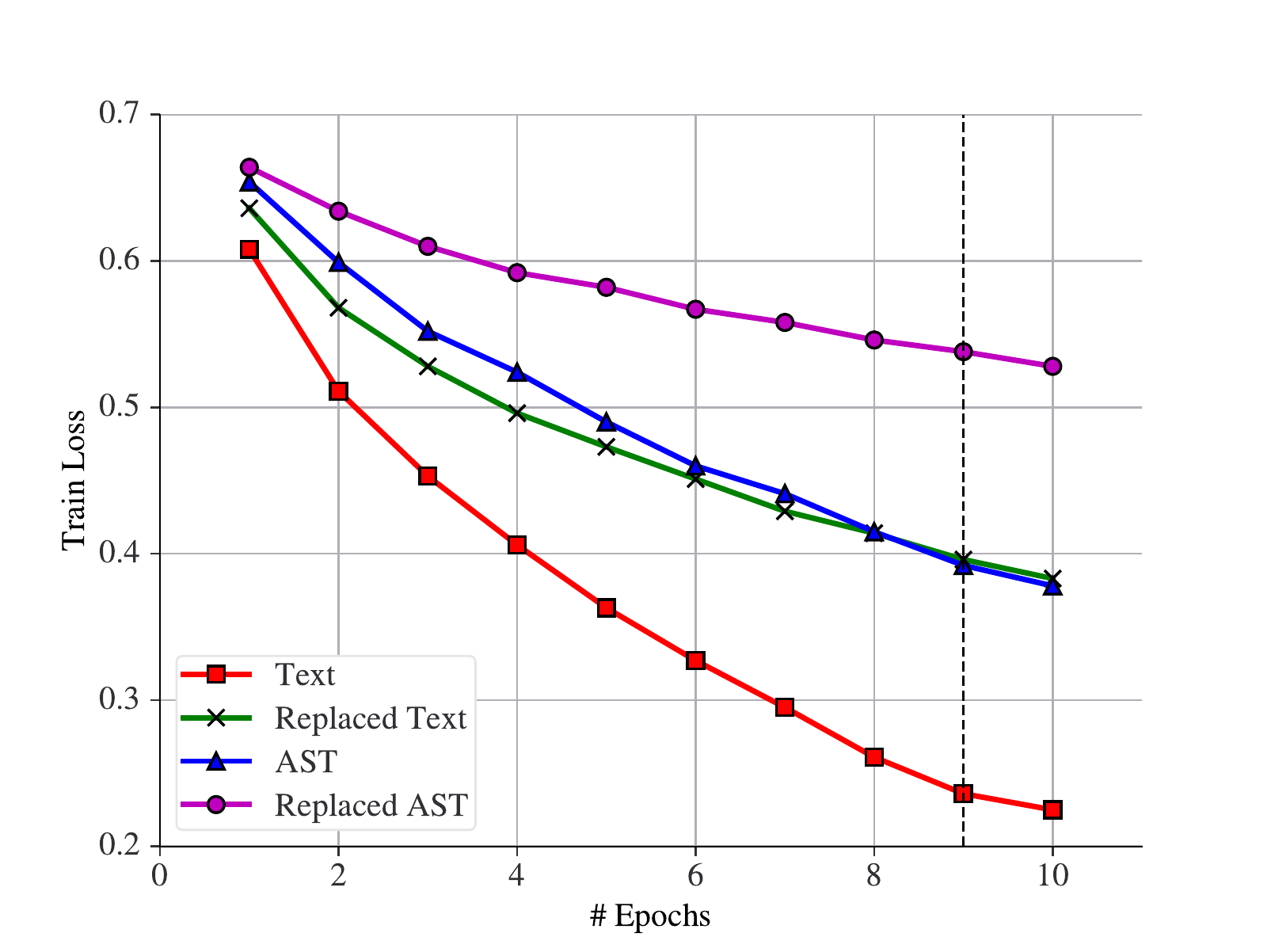}
\caption{Average loss for the training set in models trained on the four source code representations.}
\label{fig:trainloss}
\end{figure}

\begin{figure}[htbp!]
\includegraphics*[width=7.5cm]{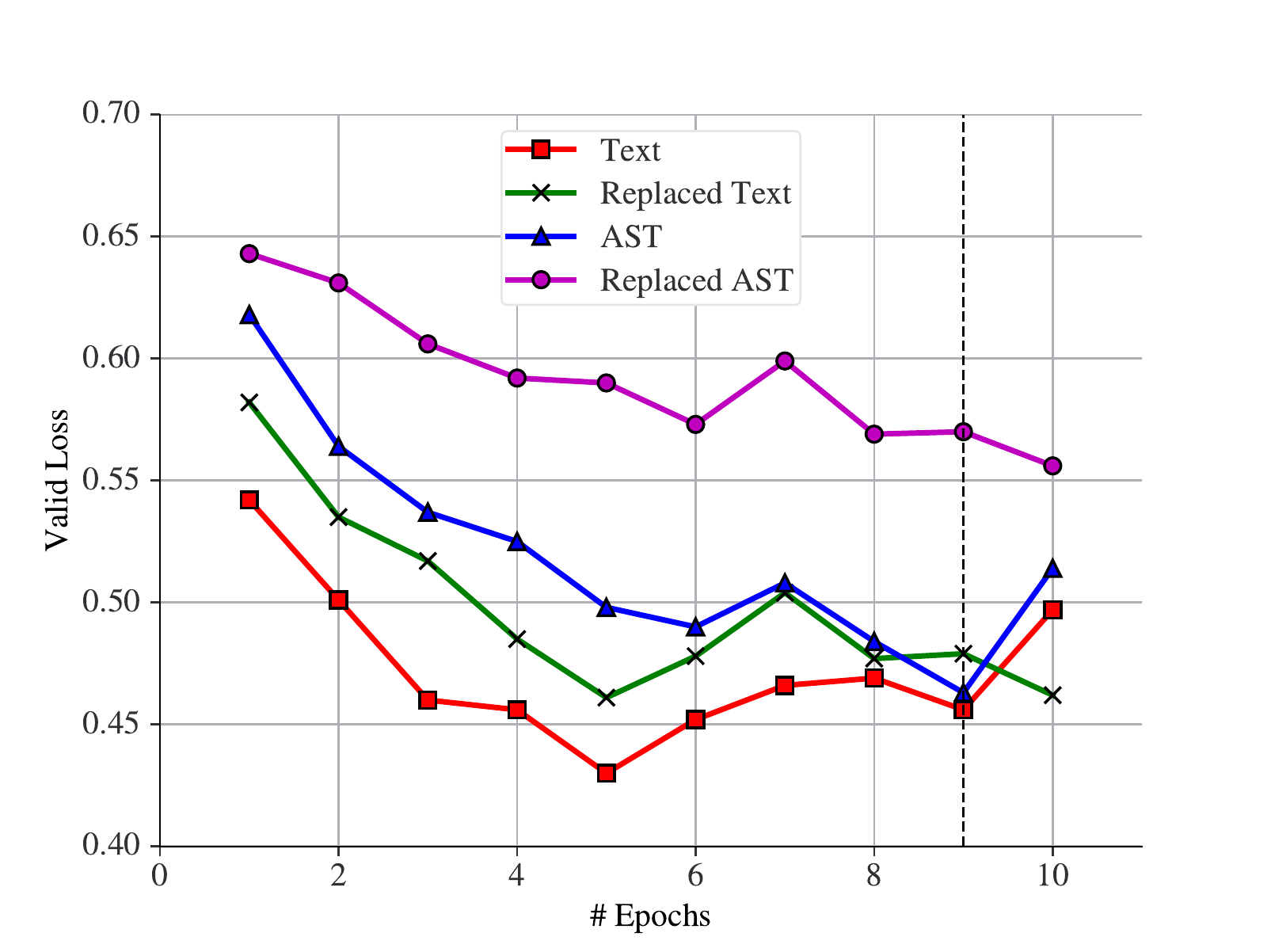}
\caption{Average loss for the validation set in models trained on the four source code representations.}
\label{fig:validloss}
\end{figure}

In \autoref{fig:accuracy} we present the accuracy scores as a function of the number of completed training epochs (passes over the full training set).
Both versions of raw text representation obtain better results than their corresponding AST representations, inconsistent with findings for other code-related tasks mentioned in~\S\ref{sec:code_representation}.
Moreover, there is a slight advantage (of about 2\%) obtained by the simple text representation over the replaced text representation.
We believe this can be explained by an implicit pattern present in parallelizable loops: they tend to have the same \say{unique} naming convention.
For example, iteration variables tend to be named \emph{i, j, k}, and \emph{A, B, C, vec, arr} as matrices and vectors.
In addition, consider a file that contains multiple OpenMP directives; in such a case, the parallelized loops are likely operating on the same variables. Therefore, the model obtains better results by recognizing these \say{unique} names.
We note that these results also appeared in our more basic BoW setup (which will be described in \S\ref{sec:classresults}), suggesting our hypothesis extends beyond the specific transformer architecture.
Nevertheless, we hypothesize that AST representations may be more useful for models whose pre-training step includes introduction of this syntax, and leave this experiment to future work. 

We provide the model's internal calculations of objective loss (Eq.~\ref{eq:cross_entropy}) over the training set and the validation set in \autoref{fig:trainloss} and \autoref{fig:validloss}, respectively.
This analysis assists us in determining the extent to which the model is \textit{overfitting}---tuning its parameters to fit spurious relations in the training set which do not generalize to the untuned-upon validation set. 
Since the validation loss curve (marked by a dashed black line) converges after 7--9 epochs, we choose to use the models trained up to those points, and report final accuracy scores of 81\% for raw text, 78\% for replaced text, 76\% for AST, and 69\% for replaced AST.
We continue our experiments with the \textsc{text} variant. 
\vspace{10ex}
\subsection{OpenMP Directive Classification}
\label{sec:classresults}

We compare \ourmod{} to two divergent methods on the task of directive classification (\textit{RQ1}): given a code snippet, is it possible to add an OpenMP pragma to parallelize it?
Our first competitor is a statistical trained Bag-of-Words (BoW)~\cite{zhang2010understanding} model with a logistic regression classifier.
In BoW, which has also been used in computer vision models~\cite{law2014bag, sikka2012exploring}, individual tokens in the input text (code) are counted and populated in a count vector containing only non-negative integers.
The order and structure of the text are not taken into account.
The classifier learns a weight vector, where each entry corresponds to a word in the vocabulary, multiplied to produce a score for each instance, traversing over a training set using gradient descent, as is standard in ML.
The second system we evaluate against is the state-of-the-art S2S compiler \compar~\cite{mosseri2020compar1,mosseri2020compar2}.
\compar{} is the most suitable for this task, as it incorporates several S2S compilers together (Par4All~\cite{par4allhome}, AutoPar~\cite{quinlan2000rose} and Cetus~\cite{cetushome}) and combines their results to produce the best OpenMP directive.
In practice, when we applied the system to our code segments, we found that only Cetus managed to compile the examples successfully.
While \compar{} generates the directives as a whole, for our current experiment we only consider the binary fact of whether or not it managed to insert an OpenMP directive into the code.

As the textual representation produced the best results for \ourmod{} during the validation setup (\S\ref{sec:validres}), it was chosen as the exclusive representation for \ourmod{} and BoW.
We trained \ourmod{} and the BoW model over the training set (\S\ref{sec:split}).
Being a deterministic compiler, \compar{} doesn't need a training step; thus, it was executed on the test set exclusively.
However, out of the 1,274 examples in the test set, \compar{} failed completely to compile 221 code instances due to complex structure definitions and operations unrecognized by its internal parser, emphasizing its limited robustness.
We thus followed a fall-back strategy that considers these cases as a \textit{negative} outcome.
Nevertheless, we report that similar results in all three metrics are obtained by conducting the same experiment with the 1,053 examples for \ourmod{}, BoW, and \compar{}. 




For evaluation, we report precision, recall, and F1 score~\cite{goutte2005probabilistic} on the test set as the performance measurements. 
These performance measurements are the standard method of evaluating classification problems: precision is the ratio between the number of possible directives predicted correctly (true positives) and the total number of directives predicted (all positives); recall is the ratio between the number of directives predicted correctly (true positives) and number of cases where directives are possible (true positives + false negatives).
F1 is the harmonic mean between precision and recall,
penalizing models which do not balance well between the two measures. 

\begin{table}[H]
\begin{tabular}{lrrrrr}
\toprule
   & \textbf{Precision}  & \textbf{Recall}  & \textbf{F1}  & \textbf{Accuracy}\\ 

\midrule
\textbf{\ourmod} & \textbf{0.80} & \textbf{0.81} & \textbf{0.80} & \textbf{0.80}
 \\
  BoW + Logistic & 0.73 & 0.74 & 0.73 & 0.74
 \\
  \compar{} &  0.51   &   0.56   &   0.36 & 0.5
 \\
\bottomrule
\end{tabular} 
\caption{Comparison between \ourmod{} and the competing systems on the task of identifying the need for an OpenMP directive.}
\label{table:directive_predict}
\end{table}

In \autoref{table:directive_predict} we present the performance of \ourmod{}, BoW and \compar{} on the directive classification task.
The best results are achieved by \ourmod{} according to all measurements.
We find the main reason behind \compar's poor performance in detecting segments that can be parallelized to be the lack of association of functions, macros, and structure definitions and implementations in the code segments, a common situation for large code projects.
In addition, we found that in loops with a low iteration count, Cetus didn't insert an OpenMP directive, although the example did contain an OpenMP directive---the benefit of an OpenMP directive in such cases is presented in~\S\ref{sec:openmp_directive_classification}.
\begin{figure}[htbp!]
\begin{tikzpicture}
  \begin{axis}[,
    xbar,
    width=\axisdefaultwidth,
    height=3.5cm,
    xlabel={Error rate \%},
    ylabel=Length,
    y axis line style = { opacity = 0 },
    x axis line style = { opacity = 0 },
    tickwidth         = 0pt,
    enlarge y limits  = 0.2,
    enlarge x limits  = 0.02,
    symbolic y coords = {10, 20, 30, 40},
    nodes near coords,
  ]
  \addplot coordinates { (11,10)         (6,20)
                         (2,30)  (0.4,40) };
  \end{axis}
\end{tikzpicture}
\caption{Prediction error rate by example length}
\label{fig:histogram_failed}
\end{figure}
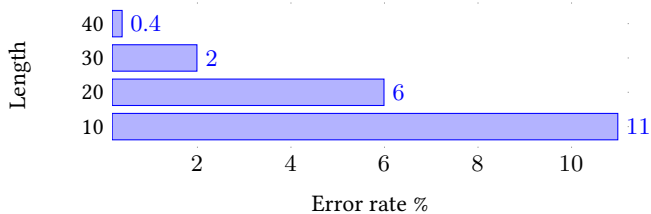
\autoref{fig:histogram_failed} presents the error rate as a function of the code's length. As seen, more than 80\% of the incorrect predictions (errors) of \ourmod{} occurred for code's with a length lower than 20. Moreover, only 10 examples with a length of more than 50 were predicted as incorrect. This might indicate that the length does not necessarily affect the decision of the model. In ~\S\ref{sec:explainability} we provide several examples showing that the attention mechanism of the model focuses on variables, function names and statements rather than other factors such as line count.

    

\subsection{OpenMP Clause Classification}
Similar to the OpenMP directive classification task, in classifying \textit{private} and \textit{reduction} clauses (\textit{RQ2}), we compared \ourmod{} to BoW and \compar{} on the same three performance metrics, this time over the \textit{clause} dataset (\S\ref{sec:split}) with balanced labels. 
We note that in a real-world application of the compared systems, directive classification will take place before clause identification, cascading errors from the previous step to this one, making our separated evaluation setup beneficial to the models which performed worse than \ourmod{} in the directive task.

We present the results for identifying the need for a \textit{private} clause in \autoref{table:private_predict} and the need for a \emph{reduction} clause in \autoref{table:reduction_predict}.
\compar{} does not perform well on precision in predicting \textit{private} clauses mostly due to the default behavior of this clause, calling for its application to the iteration variable while applying the \emph{shared} clause to the rest of the variables.
As a result, most the developers do not explicitly insert a \emph{private(i)} statement, while \compar{} does.
Having said that, \compar{} also scores low on recall, meaning that the model misses many cases where a \emph{private} clause is possible.

On the \textit{reduction} clause classification task, \compar{} manages to obtain a high score on the precision measurement, meaning that most of its positive predictions were correct.
This indicates that the conservative deterministic nature of \compar{} produces correct directives \textbf{if} it manages to generate a directive at all.
In contrast to the high precision score, and similar to its recall performance in the \textit{private} task, \compar{} scores low on recall, meaning that the model misses many cases where a \textit{reduction} clause is possible.
In contrast, \ourmod{} produces both excellent recall and precision for both clauses, demonstrating a good balance between finding many true cases without allowing many false predictions to sift through.
\begin{table}[H]
    \begin{tabular}{lrrrrr}
    \toprule
       & \textbf{Precision}  & \textbf{Recall}  & \textbf{F1}  & \textbf{Accuracy}\\ 
    \midrule
    \textbf{\ourmod }& \textbf{0.86}  & \textbf{0.85} & \textbf{0.86}   &  \textbf{0.85}
     \\
    BoW + Logistic & 0.79 & 0.78 & 0.78 & 0.79
     \\
    \compar & 0.56   &   0.51   &   0.40 & 0.56
     \\
    \bottomrule
    \end{tabular} 
    \caption{Performance of models on identifying the need for a \textit{private} clause.}
    \label{table:private_predict}
\end{table}
\begin{table}[H]
\begin{tabular}{lrrrrr}
\toprule
   & \textbf{Precision}  & \textbf{Recall}  & \textbf{F1}  & \textbf{Accuracy} \\ 
\midrule
\textbf{\ourmod} & \textbf{0.89} & \textbf{0.87} & \textbf{0.87} & \textbf{0.87}
 \\
BoW + Logistic & 0.78 & 0.78 & 0.77 & 0.78
 \\
\compar{} & 0.92   &   0.52   &   0.46 & 0.79
 \\
\bottomrule
\end{tabular} 
\caption{Performance of models on identifying the need for a \textit{reduction} clause.}
\label{table:reduction_predict}
\end{table}

\subsection{Benchmarks and Explainability}
In order to test the generality of \ourmod{}, we apply it to two existing dedicated OpenMP benchmarks that do not appear in \ourdb{}: PolyBench~\cite{polybench, polybench_git} and the Standard Performance Evaluation Corporation (SPEC)~\cite{spec_order, juckeland2014specaccel}.
The PolyBench test suite contains a collection of 30 compute-intensive benchmarks and is used, among other things, as a shared-memory environment benchmark~\cite{grauer2012autopolybench, pouchet2012polybench}.
SPEC contains several benchmarks designed to evaluate the performance of shared-memory architectures.
Among the benchmarks in SPEC are SPEC-ACCEL~\cite{spec_accl_webpage} and SPEC-OMP~\cite{spec_omp}, of which we only use the former.
We opted not to evaluate on SPEC-ACCEL as it shares many similarities with the NAS parallel benchmark (such as CG, BT, EP test-suites) already included in \ourdb{}.
PolyBench contains 64 snippets of code with OpenMP directives and 83 without.
SPEC-OMP contains 113 snippets of code with OpenMP and 174 without~\footnote{Sources: \textcolor{blue}{\url{https://github.com/pragformer/PragFormer/blob/main/DB_TEST.tar.gz}}}. \autoref{table:NAS_prediction} presents the results of \ourmod{} and \compar{} on PolyBench and SPEC-OMP. However, \compar{} failed to parse 287 snippets from the SPEC-OMP benchmark mainly due to unrecognized keywords, such as \emph{register}. Thus, we exclude them from \compar{}'s results.  The results of \ourmod{} are comparable to, and even slightly better than, the ones over the \ourdb{} test set.

\begin{table}[H]
\begin{tabular}{lrrrrr}
\toprule
   & \textbf{Precision}  & \textbf{Recall}  & \textbf{F1}  & \textbf{Acc'}\\ 
\midrule
\textbf{\ourmod{} Poly} & \textbf{0.93} & \textbf{0.93} & \textbf{0.93} & \textbf{0.93 }
\\
\compar{} Poly & 0.43 & 0.43 & 0.43 & 0.43 
\\
\textbf{\ourmod{} SPEC-OMP} & \textbf{0.81} & \textbf{0.80} & \textbf{0.80} & \textbf{0.80}
 \\
 \compar{} SPEC-OMP & 0.76 & 0.75 & 0.74 & 0.75
 \\
\bottomrule
\end{tabular} 
\caption{The performance score of \ourmod{} and \compar{} on identifying the need for an OpenMP directive on the PolyBench and SPEC-OMP benchmarks.}
\label{table:NAS_prediction}
\end{table}

\begin{table*}
  \begin{tabular}{l c c }
    \toprule
     Example & Directive & \ourmod{} prediction \\
     
    \midrule
\begin{minipage}[t]{0.5\textwidth}
\begin{minted}[linenos]{c}
for (i = 0; i < POLYBENCH_LOOP_BOUND(4000, n); i++)
  for (j = 0; j < POLYBENCH_LOOP_BOUND(4000, n); j++)
    x1[i] = x1[i] + (A[i][j] * y_1[j]);
\end{minted}
\end{minipage}
&
\begin{minipage}[t]{0.25\textwidth}
\begin{minted}{c}
#pragma omp parallel for\\
private(j)
\end{minted}
\end{minipage}
&
With OpenMP 
\\
\midrule

\begin{minipage}[t]{0.5\textwidth}
\begin{minted}[linenos]{c}
for (i = 0; i < n; i++) {
  fprintf(stderr, "%0.2lf ", x[i]);
  if ((i % 20) == 0)
    fprintf(stderr, "\n");}
\end{minted}
\end{minipage}

 &
Without OpenMP 
&
Without OpenMP 
\\
\midrule
%
\begin{minipage}[t]{0.5\textwidth}
\begin{minted}[linenos]{c}
for (i = 0; i < ((ssize_t) image->colors); i++)
  image->colormap[i].opacity = (IndexPacket) i;
\end{minted}
\end{minipage}

 &
\begin{minipage}[t]{0.25\textwidth}
\begin{minted}{c}
#pragma omp parallel for\\
schedule(dynamic,4)

\end{minted}
\end{minipage}
& 
Without OpenMP 
\\
\midrule
\begin{minipage}[t]{0.5\textwidth}
\begin{minted}[linenos]{c}
for (i = 0; i < maxgrid; i++)
  for (j = 0; j < maxgrid; j++){
    sum_tang[i][j] = (int) ((i + 1) * (j + 1));
    mean[i][j] = (((int) i) - j) / maxgrid;
    path[i][j] = (((int) i) * (j - 1)) / maxgrid; }

\end{minted}
\end{minipage}

 &
Without OpenMP
& 
With OpenMP 
\\
\addlinespace

    \bottomrule
  \end{tabular}
    \caption{Classification examples and \ourmod{} prediction.}
  \label{table:classification_examples}

\end{table*}

\begin{figure*}[htbp!]
\includegraphics*[width=\linewidth]{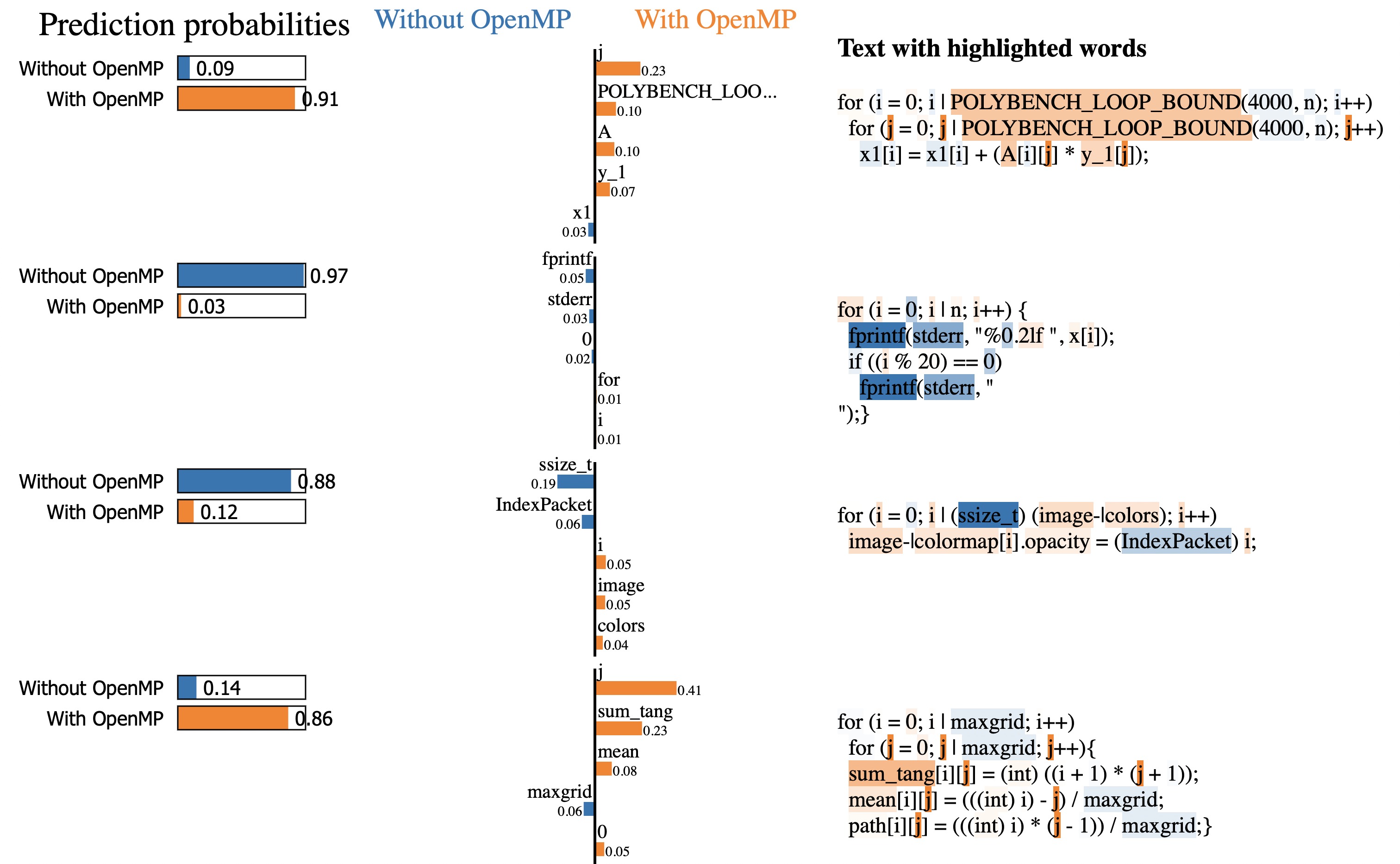}
\caption{\lime{}'s output on the four examples in \autoref{table:classification_examples}. From left to right, \lime{} presents the label distributions given to the instance by the model, followed by the most influential tokens and their weighted importance on the decision made, next highlighting their location in the input text.}
\label{fig:lime}
\end{figure*}

\label{sec:explainability}
In \autoref{table:classification_examples} we present representative examples and the corresponding prediction of \ourmod{} over the benchmark tests. Explaining and understanding the reason behind a model's prediction is a difficult task. Nonetheless, there are many algorithms that attempt to give an explanation or an intuition for classifiers' decisions \cite{adadi2018peeking}, such as \lime{} \cite{lime}. \lime{} studies the connections between keywords (tokens) of an input and the change in the prediction of a model once they are removed. Finally, \lime{} presents the probability that the keyword affected the prediction. In our case, this might indicate how \ourmod{} focuses on keywords and statements. Thus, in order to gain intuition for the predicted outcome of \ourmod{}, we applied \lime{} on the four examples 
in \autoref{table:classification_examples}. The output of \lime{} is presented in \autoref{fig:lime}.

The first example presents a code snippet taken from PolyBench, in which \ourmod{} managed to identify correctly the need for an OpenMP directive. In this example, \lime{} pinpoints the variables \emph{j}, \emph{POLYBENCH\_LOOP\_BOUND}, \emph{A} and \emph{y\_1} as the main contributors to the decision of the model. This indicates that \ourmod{} focuses on the loop variable and the arrays, as it should. The second example presents a snippet taken from PolyBench and contains an I/O operation, thus, there is no OpenMP directive. \lime{} identifies the keyword \emph{fprintf} and \emph{stderr} as the reason why the model classified the snippet without OpenMP. To verify this claim, we removed these two keywords, and \ourmod{} predicts an OpenMP directive. This probably indicates that \ourmod{} understands that these specific keywords are why there is no need for an OpenMP directive. The third example, taken from SPEC-OMP, contains an OpenMP directive that \ourmod{} fails to predict correctly. By observing \lime{}'s result, the two variables, \emph{ssize\_t} and \emph{IndexPacket}, are the main reason to its incorrect prediction. After removing both variables, \ourmod{} predicts an OpenMP directive. This might be due to the model's unfamiliarity with these keywords and how they affect the code. The fourth example, taken from PolyBench, contains an assignment into three arrays. While \ourmod{} predicts (correctly) that there should be an OpenMP directive, the example does not. As with the first example, \lime{} pinpoints the loop-variable \emph{j} and the arrays as the reason for its prediction. This further indicates that the model does focus on the correct variables while making its decision.

In summary, it appears most likely that \ourmod{} focuses on the core variables and statements in order to predict its directive. However, there are still cases in which it fails to predict correctly, possibly due to unfamiliar keywords or statements.


\section{Conclusion \& Future Work} \label{sec:conc_fw}
In this work, we presented a novel data-driven model, named \ourmod, that aims to classify the need of an OpenMP directive and OpenMP clauses.
We also created a corpus of OpenMP code snippets which is, as far as we know, the first of its nature.
The corpus contains more than 17,000 code snippets with reliable labels of whether OpenMP directives should or should not be assigned to them. 
We invite the community to use our corpus for analyzing parallelization use-cases and as a development bed for ML models learning to parallelize code.

We trained \ourmod{} model variants which ingest different representations of code, finding that representing the source code as raw text obtains the best results.
We then found that \ourmod{} manages to perform fundamental tasks on the roadmap to parallelization better than two competitors---a top-shelf deterministic S2S compiler and a lightweight text-aware ML model. 

In most cases, the context of the loop itself is enough to generate an OpenMP directive. However, in order to push performance further, it is effective to include context from previous and next lines of code. Thus, we will expand the database that it will contain these extra lines of code.
In addition, in order to improve the effectiveness of \ourmod, we plan to explore additional code representations such as the structured based traversal (SBT) representation~\cite{hu2018deep, hu2020deep, leclair2020improved, li2022setransformer}, combination of text and AST, \emph{IR2vec}, \emph{Programl} and more. 
We also plan to expand the corpus by searching other keywords on Github and other resources, as well as including code snippets from C++ and Fortran as well as C. 
Most importantly, \ourmod{} provides a crucial step towards a more robust framework based on NLP techniques that will eventually be able to insert OpenMP directives automatically with high fidelity, contributing to making the world's code more efficient. 

Evaluating the effectiveness of inserting OpenMP directives automatically via \ourmod{} will be done by comparing it to the state-of-the-art S2S compiler---\compar{}, which produces the optimal OpenMP directive generated by three S2S compilers.
In addition, it can fine-tune the OpenMP directives by inserting the \textit{scheduling} construct and executing the source code to determine the best combination.
Thus, \compar{} is suitable for evaluating the directives inserted by \ourmod{} and for comparing their execution times.

Once we have proven the capacity of transformers learning to generate and identify OpenMP parallelization,
we intend to replicate this process for other shared memory parallelization schemes for more challenging frameworks such as CUDA and OpenCL.
In addition, we can explore other parallelization schemes such as the MPI scheme for distributed memory~\cite{openmpi4}.
Finally, combining all the models may result in a framework capable of producing an \emph{MPI+X} parallelization scheme. 

\begin{acks}
This research was partially supported by the Israeli Council for Higher Education (CHE) via the Data Science Research Center, Ben-Gurion University of the Negev, Israel, and 
the Lynn and William Frankel Center for Computer Science.
Computational support was provided by the NegevHPC project~\cite{negevhpc}. 
\end{acks}



\bibliographystyle{ACM-Reference-Format}
\bibliography{sample-base}

\end{document}